\documentclass[traditabstrac]{aa} % for the letters 
\newcommand{\earth}{\oplus}
\newcommand{\micron}{$\mu$m}

\newcommand{\kms}{km\,s$^{-1}$}
\usepackage{graphicx}
\usepackage{txfonts}

\begin{document}

   \title{ALMA unveils rings and gaps in the protoplanetary system \\
    HD 169142: signatures of two giant protoplanets }

   \author{
   D.\ Fedele\inst{\ref{inst_inaf}}, 
   M.\ Carney\inst{\ref{inst_leiden}},
   M. R. \ Hogerheijde\inst{\ref{inst_leiden}}
   C. \ Walsh\inst{\ref{inst_leiden}, \ref{inst_leeds}},
   A. \ Miotello\inst{\ref{inst_leiden}},
   P. \ Klaassen\inst{\ref{inst_roe}},
   S. \ Bruderer\inst{\ref{inst_mpe}},\\
   Th. \ Henning\inst{\ref{inst_mpia}},
   E.F. \ van Dishoeck\inst{\ref{inst_leiden}, \ref{inst_mpe}}   }
                 
\institute{
INAF-Osservatorio Astrofisico di Arcetri, L.go E. Fermi 5, I-50125 Firenze, Italy\label{inst_inaf}\\ 
\email{fedele@arcetri.astro.it}
\and
Leiden Observatory, Leiden University, P.O. Box 9513, NL-2300 RA, Leiden, The Netherlands\label{inst_leiden}\\
\and
School of Physics and Astronomy, University of Leeds, Leeds LS2 9JT, UK\label{inst_leeds}\\
\and
UK Astronomy Technology Centre, Royal Observatory Edinburgh, Blackford Hill, Edinburgh EH9 3HJ, UK\label{inst_roe}\\
\and
Max Planck Institut f\"{u}r Extraterrestrische Physik, Giessenbachstrasse 1, 85748 Garching, Germany\label{inst_mpe}\\
\and
Max-Planck-Institute for Astronomy, Koenigstuhl 17, 69117 Heidelberg, Germany\label{inst_mpia}\\
}

\authorrunning{
   D.\ Fedele\ et al.}
   \date{}

% \abstract{}{}{}{}{} 
% 5 {} token are mandatory
 
  \abstract
   {
   The protoplanetary system HD 169142 is one of the few cases where a potential candidate protoplanet has been recently detected via direct imaging in the 
   near-infrared. To study the interaction between the protoplanet and the disk itself observations of the gas and dust surface density structure are needed.
   This paper reports new ALMA observations of the dust continuum at 1.3\,mm,  $^{12}$CO, $^{13}$CO and C$^{18}$O $J=2-1$ emission from the (
   almost face-on) system HD 169142 at angular resolution of $\sim 0\farcs3 \times 0\farcs2$ ($\sim 35 \times 20\,$au).  
   The dust continuum emission reveals a double-ring structure with an inner ring between $0\farcs17-0\farcs28$ ($\sim 20 - 35\,$au) and an outer ring 
   between $0\farcs48-0\farcs64$ ($\sim 56 - 83\,$au). The size and position of the inner ring is in good agreement with previous polarimetric observations 
   in the near-infrared and is consistent with dust trapping by 
   a massive planet
   No dust emission is detected inside the inner dust cavity ($R \lesssim 20\,$au) or within the dust gap ($\sim 35 - 56\,$au) down to the noise level. 
   In contrast, the channel maps 
   of the $J=2-1$ line of the three CO isotopologues reveal the presence of gas inside the dust cavity and dust gap. 
   The gaseous disk is also much 
   larger than the compact dust emission extending to $\sim 1\farcs5$ ($\sim 180\,$au) in radius. This difference and the sharp drop of the continuum 
   emission at large radii point to radial drift of large dust grains ($>$ \micron-size). Using the thermo-chemical 
   disk code \textsc{dali}, the continuum and the CO isotopologues emission are modelled to quantitatively measure the gas and dust surface densities. The resulting gas
   surface density is reduced by a factor of $\sim 30-40$ inward of the dust gap. The gas and dust distribution hint at the presence of 
   multiple planets shaping the disk structure via dynamical clearing (dust cavity and gap) and dust trapping (double ring dust distribution).
   }

   \keywords{protoplanetary disks -- giant planet formation }

   \maketitle

\section{Introduction}
Observations of the cold gas and dust reservoirs of protoplanetary disks are powerful tools for investigating the early phase of planet formation. 
In particular, addressing the radial distribution of gas and dust in the disk interior has the potential to unveil the initial conditions for the formation 
of gas giant planets. In contrast with optical/near-infrared scattered light data or infrared spectroscopy, sub-millimetre observations probe the bulk of the gas and dust mass in the disk.
Recent observations in the sub-millimetre regime with the Atacama Large Millimetre/Sub-millimeter Array (ALMA) brought to light the evidence of discontinuous radial (gaps and cavities) and azimuthal (traps and vortices) dust distributions \citep[e.g.,][]{Casassus12, vdMarel13, ALMA15, Walsh14, Andrews16, Isella16} and cold 
gas cavities that are smaller than the dust cavities \citep{vdMarel16}. 

\noindent
Different theories have been proposed to explain the formation of dust gaps and rings including: 
dynamical interaction with one or more giant planets carving out the dust \citep[e.g.,][]{Papaloizou84}; magneto-rotational instability giving rise to 
dead-zones \citep{Flock15}; dust grain growth corresponding to condensation fronts \citep{Zhang15}; fusing of dust grains at temperatures below the sublimation point  \citep[dust sintering,][]{Okuzumi16}; 
photoevaporation \citep{Ercolano17}.

\noindent
Sub-millimetre interferometric observations of both gas and dust have the potential to unveil the evolutionary status of protoplanetary systems and to 
disentangle the physical origin of dust cavities and dust gaps in disks. 
A particularly powerful test is to perform ALMA observations of disks for which optical/near-infrared data have previously suggested embedded planets.

\begin{table*}[!htbp]
 \caption{HD 169142 observational parameters}
 \centering
 \label{tab:obs_par}
 \begin{tabular}{lccc}
 \hline \hline
 Dates Observed & \multicolumn{3}{c}{2015 August 30} \\
 Baselines & \multicolumn{3}{c}{13 -- 1445 m | 10 -- 1120 k${\rm \lambda}$ } \\
 \hline
  & C$^{18}$O 2--1 & $^{13}$CO 2--1 & $^{12}$CO 2--1 \\
 Rest frequency [GHz]  & 219.56035 & 220.39868 & 230.53800  \\
 Synthesized beam [FWHM] &  $0\farcs36 \times 0\farcs23$ & $0\farcs37 \times0\farcs22 $ & $0\farcs37 \times0\farcs20$ \\
 Position angle &  --74\fdg5 & --75\fdg2 & --72\fdg8 \\
 Channel width [km s$^{-1}$] & 0.084 & 0.084 & 0.040 \\
 r.m.s. noise (per channel) [mJy beam$^{-1}$]  & 6 & 8 & 13 \\
Peak emission [mJy beam$^{-1}$] & 100 & 200 & 540 \\
 Integrated flux\tablefootmark{a} [Jy km s$^{-1}$]  & 3.9$\pm$0.5 & 7.6$\pm$0.6 & 14.0$\pm$2.0 \\
 Weighting & \multicolumn{3}{c}{natural} \\
 \hline
 Continuum Frequency [GHz] & \multicolumn{3}{c}{233.0}  \\
 Synthesized beam [FWHM] &  0\farcs28 $\times$ 0\farcs18 & 0\farcs24 $\times$ 0\farcs16 & 0\farcs22 $\times$ 0\farcs14  \\
 Position angle &  --77\fdg9 &  --78\fdg3 &  --80\fdg0 \\
 r.m.s. noise [mJy beam$^{-1}$]  & 0.07 & 0.26 & 0.11 \\
 Peak emission [mJy beam$^{-1}$] & 17 & 15 & 13  \\
 Integrated flux [mJy]  & 232$\pm$23 & 226$\pm$23  & 226$\pm$23   \\
 Weighting & Briggs, robust = 0.5  & Uniform  & Superuniform \\
 \hline
 \end{tabular}
 \tablefoot{Flux calibration accuracy is taken to be 10\%. 
 \tablefoottext{a}{Integrated over a circular aperture of 3\arcsec \ radius.}
 }
\end{table*}

\section{HD 169142}
HD 169142 is a young $6^{+6}_{-3}$\,Myr and isolated \citep{Grady07} Herbig Ae/Be star \citep{The94} with $M_{\star} = 1.65\,M_{\odot}$ 
\citep{Blondel06}, spectral type A5 and $T_{\rm eff} = 8400\,$K \citep{Dunkin97}. The most recent measurement of the parallax is 
$\varpi = 8.53 \pm 0.29$\,mas \citep{Brown16} which translates to a distance $d = 117 \pm 4\,$pc. Previous estimates by \citet{deZeeuw99} 
had instead $d = 145\,$pc. In the rest of the paper we will use the newest distance estimate adjusting all the relevant parameters.
The stellar luminosity adopted in this paper is $L_{\star} = 10\,L_{\odot}$ based on the new distance estimate and on the the optical V-magnitude 
and extinction (V = 8.15\,mag, $A_{\rm V}=0.43\,$mag, \citealt[e.g.,][]{Malfait98a}).
Sub-millimeter observations of the 1.3\,mm dust continuum and CO $J=2-1$ with the Submillimeter Array (SMA) measured a disk inclination of 
13$^{\circ}$ and a position angle of 5$^{\circ}$  \citep{Raman06, Panic08}. Based on observations of multiple CO isotopologues with the SMA, 
\citet{Panic08} derive a total gas mass of $0.6-3.0 \times 10^{-2}\,M_{\odot}$ in good agreement with the estimate by \citet{Pinte10} of 
$\sim 10^{-2}\,M_{\odot}$ based on the {\it Herschel}/PACS detection of [\ion{O}{I}] 63\,\micron ~ \citep{Meeus10, Fedele13a}. 
The spectral energy distribution (SED) suggests the presence of a discontinuous radial distribution of the dust \citep{Malfait98a}. 
This is confirmed by direct imaging observations of the thermal emission at mid-infrared wavelengths \citep{Honda12} as well as by $H-$band 
scattered light emission \citep{Fukagawa10, Quanz13, Momose15, Wagner15}. In particular,  the H-band polarimetric image shows a ring-like 
dust distribution at a radius of $\sim 0\farcs17$ (20\,au) from the star.  The dust ring seen in scattered light is also detected in the 7\,mm 
continuum \citep{Osorio14}.

\smallskip
\noindent
\citet{Biller14} and \citet{Reggiani14} detected a point-like emission at $\sim 0\farcs11 - 0\farcs16$ ($13-18\,$au) via $L'$-band coronographic and polarimetric differential imaging, respectively. According to \citet{Reggiani14} this emission is produced by a massive planet of  $\sim 30\,M_{\rm Jupiter}$. 
\citet{Biller14} warn instead that, because of the non-detection in the $H$ and $K_{\rm s}$ bands, this should be a disk feature.
Nevertheless, the potential discovery of a protoplanet makes HD 169142 an ideal case to study the planet-disk interaction during the early phases of planetary formation and evolution. 

\smallskip
\noindent
This paper presents new ALMA high angular resolution observations of HD 169142 of the $J=2-1$ transition of $^{12}$CO, $^{13}$CO and C$^{18}$O and 1.3\,mm dust continuum.
From these observations we place more stringent constraints on the dust and gas density structures. In section 3 we summarise the observing strategy and the data reduction. Results
are presented in section 4. In section 5 we compare the observations with simulations of thermo-chemical disk models to constrain the gas and dust distributions. Discussion and 
conclusions are given in sections 6 and 7, respectively. 

\begin{figure*}
\centering
\includegraphics[width=18cm]{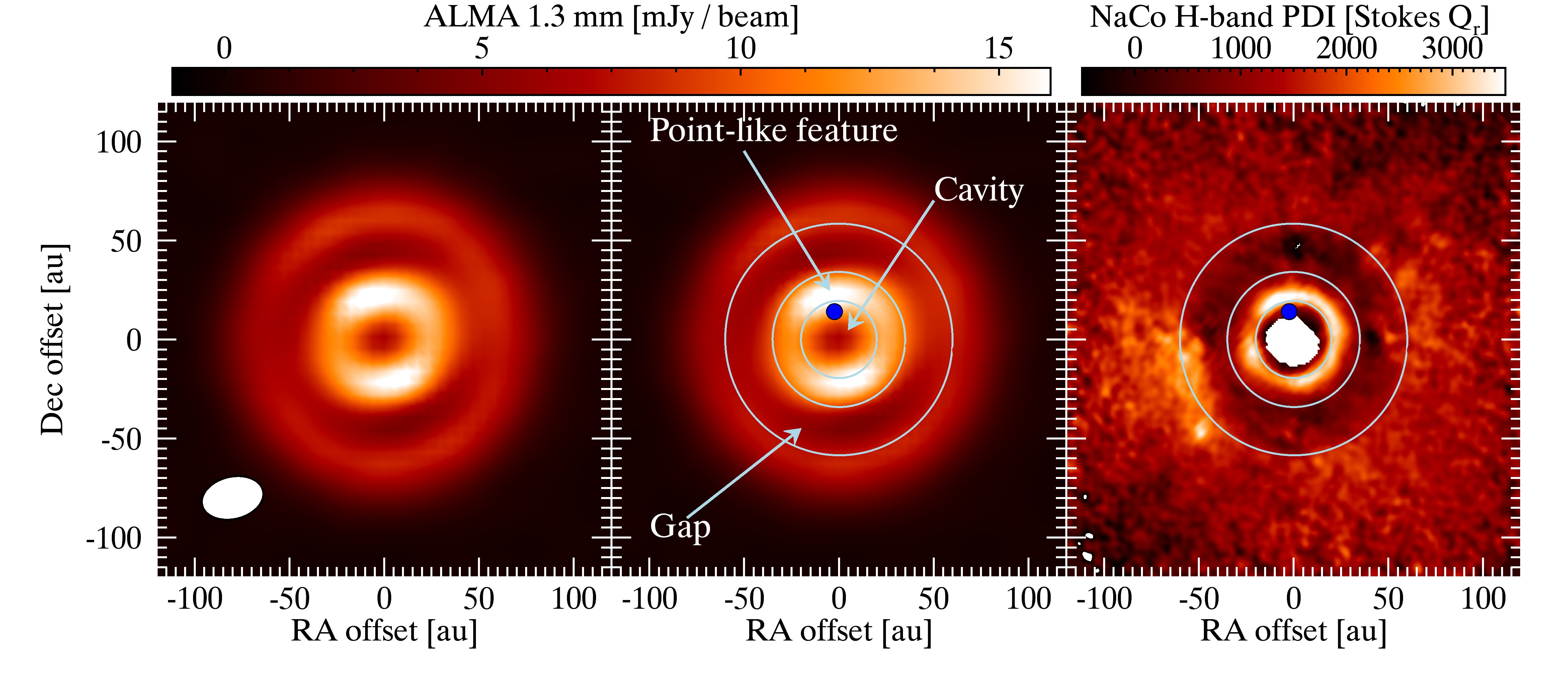}
\caption{(left) ALMA 1.3\,mm continuum map  with Briggs weighting, robust=0.5, (center) overlaid with the position and size of the inner dust cavity and gap and the position of the $L'$-band point like feature. (right) NaCo H-band polarimetric differential image \citep{Quanz13}.}\label{fig:cont}
\end{figure*}

\begin{figure*}
\centering
\includegraphics[width=18cm]{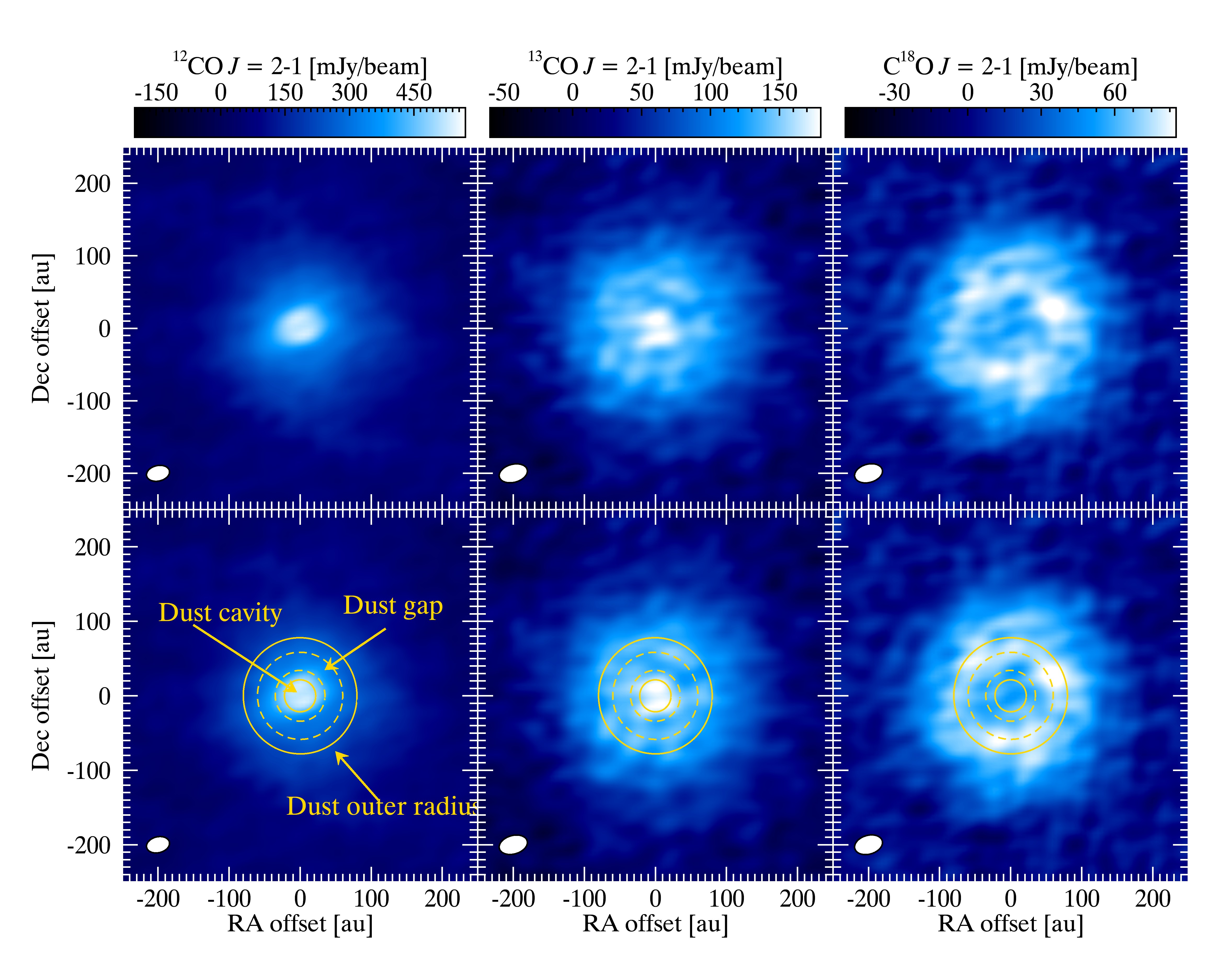}
\caption{(top) Integrated intensity maps (natural weighting) of  $^{12}$CO (left), $^{13}$CO (center) and C$^{18}$O (right) $J=2-1$, (bottom) 
and overlaid with the dust rings structure. }\label{fig:lines}
\end{figure*}

\begin{figure}
\includegraphics[width=9cm]{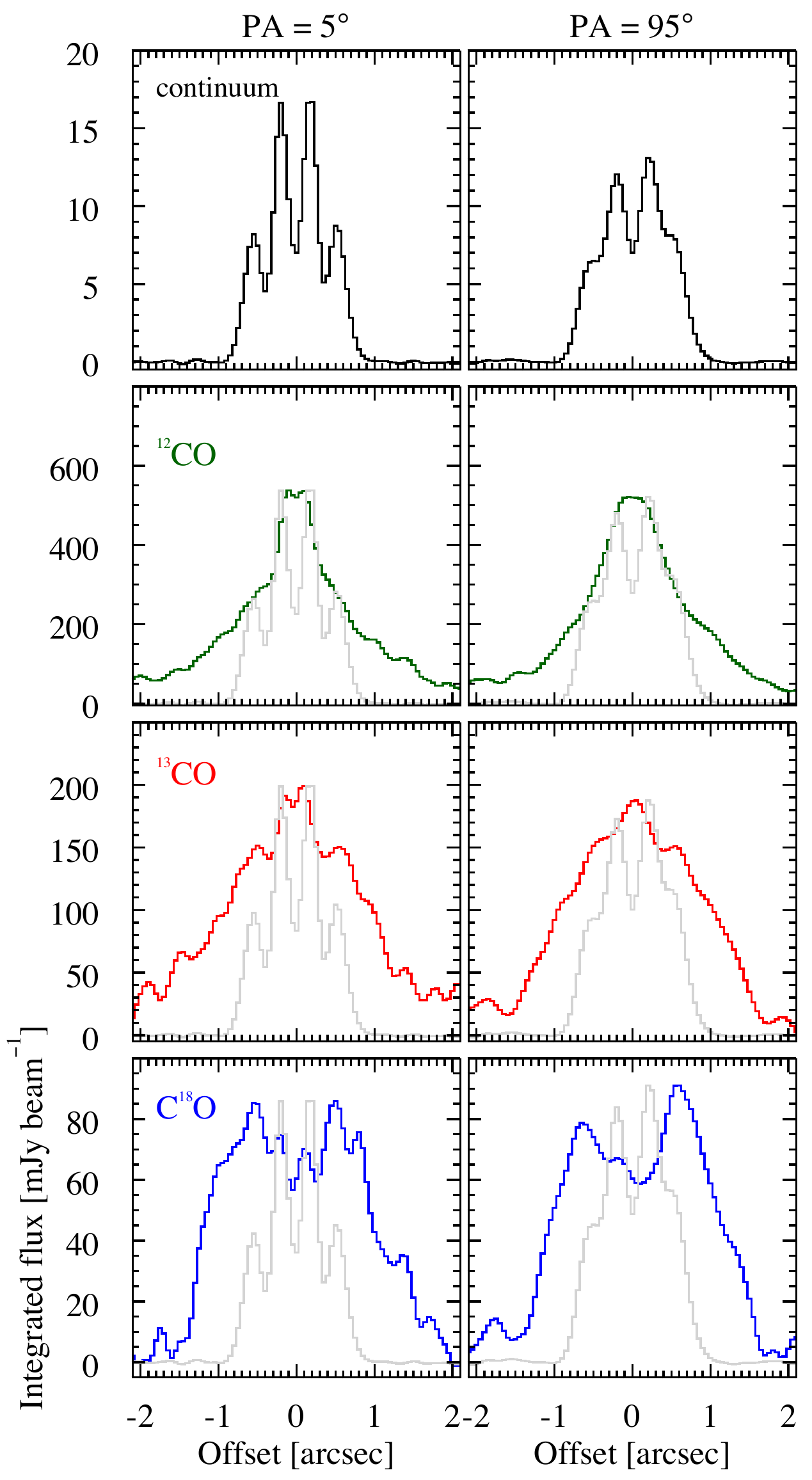}
\caption{Intensity profiles of the dust continuum and CO isotopologues integrated emission maps along the major (PA=5$^{\circ}$, left) and minor
(PA=95$^{\circ}$, right) disk axis.  The normalized continuum profile is overlaid (grey) on the CO panels for comparison. Note that the large scale 
($> 2\arcsec$) emission of $^{12}$CO and $^{13}$CO is the result of the reduced image quality (see Sec.~2).}\label{fig:radcut}
\end{figure}

\begin{figure}
\centering
\includegraphics[width=8cm]{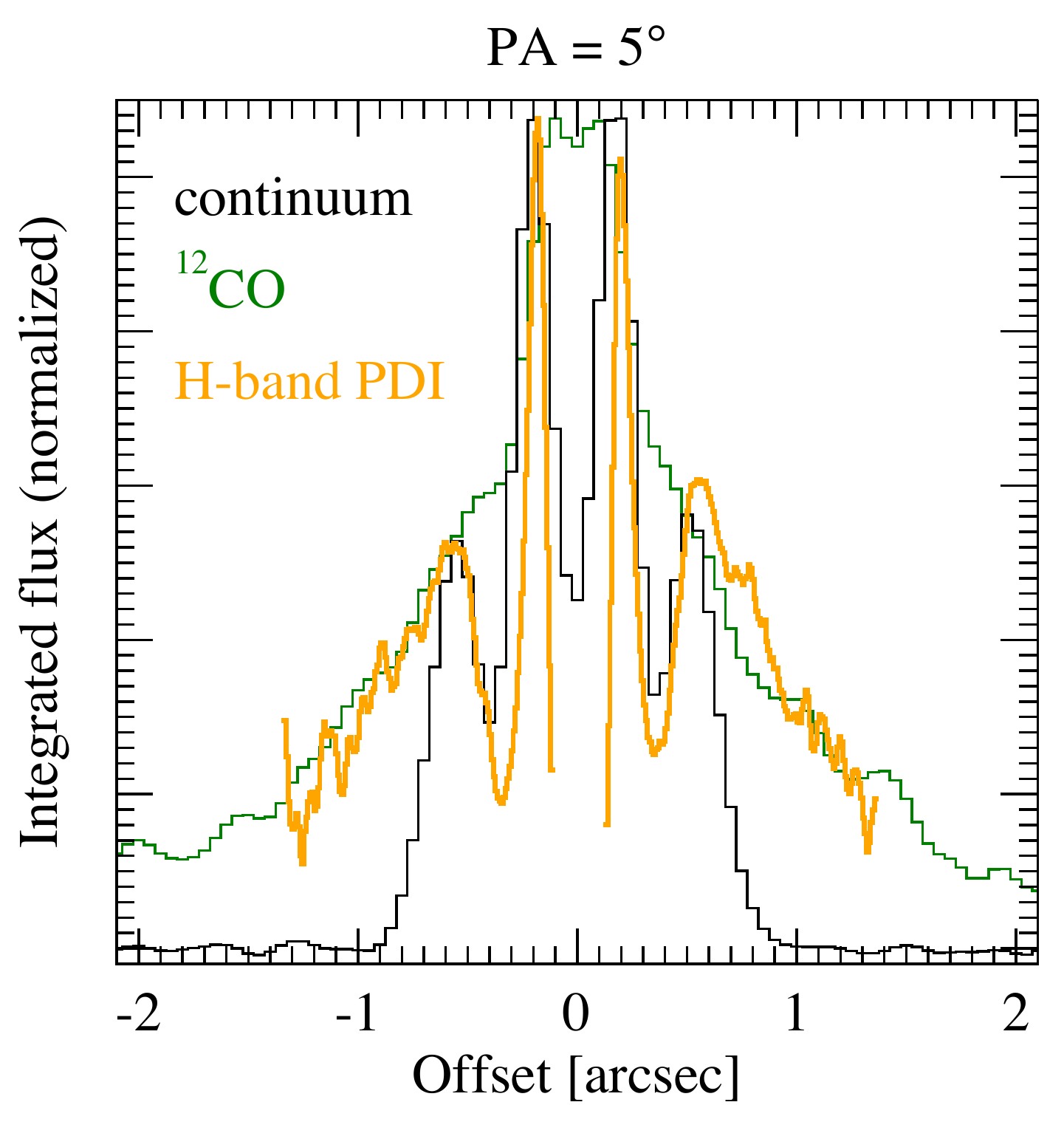}
\caption{Intensity profile of the H-band scattered light emission (azimuthally averaged to 
increase signal-to-noise) overlaid on the 1.3~mm continuum and $^{12}$CO profiles.}\label{fig:radcut_naco}
\end{figure}

\section{Observations and data reduction}
HD 169142 (J2000: R.A. = 18$^{\rm{h}}$24$^{\rm{m}}$29.776$^{\rm{s}}$, DEC = --29$^\circ$46$\arcmin$49.900$\arcsec$) was observed on 2015 August 
30 with the Atacama Large Millimeter Array (ALMA) in band 6 (211--275\,GHz) as part of project 2013.1.00592.S. In total, 35 antennas were used  to 
achieve a spatial resolution of $\sim 0\farcs2 - 0\farcs3$. The upper sideband (USB) 
contained two spectral windows. One window has continuum observations in the Time Domain Mode (TDM) correlator setting with 2 GHz bandwidth 
centered at 233\,GHz. The $^{12}$CO $J$ = 2--1 line at 230.538\,GHz was observed in the second USB spectral window with the Frequency Domain 
Mode (FDM) correlator setting at 30.5\,kHz (0.040\,km\,s$^{-1}$) frequency (velocity) resolution. $^{13}$CO $J$ = 2--1 at 220.39868 GHz and
C$^{18}$O $J$ = 2--1 at 219.56035\,GHz were both observed in separate spectral windows in the lower 
sideband (LSB) in FDM mode. Each observed LSB line had a frequency (velocity) resolution of 61.0 kHz (0.084\,km\,s$^{-1}$). 
Table~\ref{tab:obs_par} summarizes the observational parameters. 

\smallskip
\noindent
Visibility data were obtained in a single execution block with a 6.05s integration time per visibility for 50 minutes total on-source. System 
temperatures were between $50-200$\,K. Weather conditions on the date of observation gave an average precipitable
water vapour of 1.8\,mm. Calibration was done with $J1924-2914$ as the delay and bandpass calibrator, J1812-2836 as the gain calibrator, and Ceres 
as the flux calibrator. The flux values for Ceres on the date of observation were 1.941\,Jy in the LSB and 2.165\,Jy in the USB. The visibility 
data were subsequently time binned to 30s integration times per visibility for self-calibration, imaging, and analysis.
Extended emission is present in $^{12}$CO and $^{13}$CO data and poorly sampled on short baselines in the $uv-$space, which resulted in a reduced
image quality. For this reason, the total flux listed in Table~\ref{tab:obs_par} is integrated in a circular aperture of 3\arcsec \ radius centered on the source position. 

\smallskip
\noindent
Self-calibration for HD 169142 was performed using the 233 GHz continuum TDM spectral window with DA59 as the reference antenna. Calibration 
solutions were calculated twice for phase and once for amplitude. The first phase solution interval was set to 300s, the second phase and 
amplitude solutions had the solution interval set equal to the binned integration time. Self-calibration solutions from the TDM spectral 
window were cross-applied to each FDM spectral window. Continuum subtraction of the line data was done in $uv-$space based on a single-order 
polynomial fit to all line-free channels in each spectral window. \textsc{clean} imaging for the continuum was done using different weighting schemes, Briggs \citep{Briggs95}, uniform and superuniform, (Table~\ref{tab:obs_par}). In the rest of the paper we adopt the first (which provides the minimum r.m.s.). For the lines \textsc{clean} imaging was done with natural weighting. Data reduction 
was performed with the Common Astronomy Software Applications (\textsc{casa}, \citealt{Mcmullin07}). 

\section{Results}
The 1.3\.mm continuum and all the three CO isotopologues lines are readily detected. Figures~\ref{fig:cont},  \ref{fig:lines} and \ref{fig:radcut} show the dust continuum map, the lines integrated intensity maps and the radial profiles, respectively. The channel maps of the three lines are presented in Figures~\ref{fig:channels_12co}, \ref{fig:channels_13co} and \ref{fig:channels_c18o} in the Appendix.

\subsection{Dust continuum emission}
The 1.3\,mm continuum map (Figure~\ref{fig:cont}) and the radial profile (Figure~\ref{fig:radcut}) reveal a double-ring structure in the dust distribution 
with an inner cavity $\sim$ 0\farcs17 in radius and a dust gap between $\sim 0\farcs28-0\farcs48$. The dust continuum emission drops steeply beyond 0\farcs64. The different structures are highlighted in Figure~\ref{fig:cont} (center panel) along with the position of the point-like $L'$-band emission 
\citep{Biller14, Reggiani14}. 
The radial profile (Figure~\ref{fig:radcut}) is shown at 2 different position angles, PA=5$^{\circ}$ (major axis) and PA=95$^{\circ}$ (minor axis).
Along the minor axis, the continuum is slightly asymmetric with the West side brighter than the East side. The flux difference between the two sides is
$\sim 17\,$mJy ($\sim 2.5\sigma$).

\smallskip
\noindent
The ALMA continuum map shows some similarities with the H-band polarimetric differential imaging (PDI, \citealt{Quanz13, Momose15}). 
The NaCo H-band PDI image is shown in Figure~\ref{fig:cont} and Figure~\ref{fig:radcut_naco} shows the radial intensity profile. The position
 and size of the inner dust ring and gap are consistent between both wavelength ranges. 
 In the outer disk, the ALMA continuum emission is clearly more compact than the H-band emission (Figure~\ref{fig:radcut_naco}). 

\noindent
The dearth of dust continuum emission inside the inner dust cavity and the dust gap, and the similarities between the H-band PDI and the dust 
continuum emission, suggest that the cavity and the gap are due to a substantial depletion of dust particles. 
An upper limit to the dust mass inside the gap can be estimated from the r.m.s. of the continuum flux density (Table~\ref{tab:obs_par}). With the
assumption of optically thin emission, the dust mass is \citep[e.g.,][]{Roccatagliata09}:

\begin{equation}
M_{\rm dust, gap} = \frac{S_{\rm \nu, gap} \ d^{2}}{k_{\nu} \ B_{\nu} \ (T_{\rm dust, gap})}
\end{equation}

\noindent
where $S_{\rm \nu, gap}$ (Jy) is the upper limit on the flux density, $d$ (cm) the distance, $k_{\nu}$ = 2  (cm$^{2}$ \ g$^{-1}$) the mass absorption 
coefficient at 230\,GHz \citep{Beckwith90}, $T_{\rm dust, gap}$ (K) the dust temperature inside the gap and $B_{\nu}$ (Jy sr$^{-1}$) the Planck function. 
We assume $T_{\rm dust, gap} = 50\,$K (see
 Fig.~\ref{fig:dali}). The flux density upper limit is computed adopting a constant flux of $2.1 \times 10^{-4}$ Jy beam$^{-1}$ (i.e., 3 $\times$ r.m.s.) over the
 entire gap area. This corresponds to a dust mass 3~$\sigma$ upper limit of $\sim$ 0.3 M$_{\earth}$. With the same assumptions the dust mass upper limit inside the cavity (assuming $T_{\rm dust, cavity} = 150\,$K, Fig.~\ref{fig:dali}) is $\sim 10^{-2}$ M$_{\earth}$.

\subsection{CO isotopologues emission}
The integrated intensity maps (Figure~\ref{fig:lines}) and the radial profile (Figure~\ref{fig:radcut}) of the three CO isotopologues show different 
intensity distributions: the $^{12}$CO emission is centrally concentrated with most of the line intensity originating within a $\sim$ 0\farcs20 \ radius; 
the peak of the $^{13}$CO emission corresponds to that of $^{12}$CO but with a secondary ring-like structure further out in the disk; in the case of 
C$^{18}$O, the emission map shows an inner (weak) ring centered at $\sim 0\farcs1-0\farcs2$ and a (strong) outer ring peaking at 
$\sim 0\farcs55$ with tail up to $\sim 1\farcs7$. 
The gas emission is more extended than the dust continuum emission (Figure~\ref{fig:radcut}). Moreover, the H-band scattered light emission, in the outer disk, follows the same intensity distribution as that of $^{12}$CO. 

\smallskip
\noindent 
The positions of the 2 C$^{18}$O peaks are spatially coincident with the location of the dust rings. 
Along the disk minor axis, the C$^{18}$O is slightly asymmetric with the West side brighter than the East one (similar to the continuum asymmetry).
The flux difference between the 2 peaks is $\sim 18\,$mJy \ ($\sim 3\sigma$).  The line emission maps are consistent with a disk inclination of 
13$^{\circ}$ and a position angle of the disk major axis of 5$^{\circ}$.

\noindent 
The different radial distributions of the emission from the three isotopologues is readily explained by an optical depth effect as the $J=2-1$ 
transition of the three species has different $\tau$ with $\tau ({\rm ^{12}CO}) > \tau ({\rm ^{13}CO}) > \tau ({\rm C^{18}O})$. The optically thick 
$^{12}$CO emission is mostly sensitive to the gas temperature, and as a consequence its line intensity peaks toward the central, hotter, region 
of the disk. As the optical depth goes down, the line emission is less sensitive to the gas temperature and more sensitive to the gas column 
density. This is clear when looking at the distribution of $^{13}$CO and C$^{18}$O: in the first case the emission is less peaked (compared 
to $^{12}$CO) toward the central region and it also shows a secondary peak (ring-shaped) in the outer, colder, disk. Finally, the optically thin 
C$^{18}$O emission originates mostly in the outer disk showing a clear ring-like structure. The ring-like shape seen in the  $^{13}$CO and C$^{18}$O 
emission map is spatially coincident with the outer dust ring. 

\subsection{Disk surface density}
The spatial distribution of the three isotopologues emission provides direct insight into the gas content in the disk: the strong centrally peaked 
$^{12}$CO emission indicates the presence of gas inside the dust gap and the dust cavity. On the other hand, the line intensity map of $^{13}$CO 
and in particular C$^{18}$O implies a substantial drop of the gas surface density by a factor $\delta_{\rm gas}$ of the order of $\sim$ 100 inside 
the dust gap and cavity (see section 5). 
The similar intensity profiles of the scattered light and $^{12}$CO emission, in the outer disk, is a strong indication that the small dust grains are dynamically and 
thermally coupled to the gas in the outermost layers of the disk. The intensity drop in the inner disk is also clearly seen in the individual channel 
maps shown in Figures~\ref{fig:channels_12co}, \ref{fig:channels_13co} and \ref{fig:channels_c18o} in the Appendix

\smallskip
\noindent
The significance of the asymmetric emission along the minor axis (continuum and C$^{18}$O) is low ($\lesssim 3\sigma$) and further observations are needed to 
confirm such a structure.

\section{Analysis}
In this section the ALMA observations of the 1.3\,mm continuum and of the three CO isotopologues are compared with thermo-chemical disk model 
simulations. The goal is to quantify the decrease in dust and gas in the cavity and gap identified in the images.

\subsection{Disk model description}
The simulations presented here were generated using the thermo-chemical disk code \textsc{dali} (Dust and Lines, \citealt{Bruderer12, Bruderer13}). 
In this example \textsc{dali} takes as input a $T_{\rm eff} = 8400\,$K blackbody radiation field to simulate the stellar spectrum and
a power-law surface density structure with an exponential tail

\begin{equation}
\Sigma_{\rm gas}(R) = \Sigma_{\rm c}  \ \Bigg(\frac{R}{R_{\rm c}}\Bigg)^{-\gamma} \ \exp\Bigg[ - \Bigg( \frac{R}{R_{\rm c}} \Bigg)^{2 - \gamma} \Bigg]
\end{equation}

\noindent
with $R$ the radial distance from the star, $R_{\rm c}$ critical radius, $\Sigma_{\rm c}$ gas surface density at $R=R_{\rm c}$. The dust surface density
is  $\Sigma_{\rm gas}  / \Delta_{\rm gd}$, where $\Delta_{\rm gd}$ is the gas-to-dust mass ratio.
Along the vertical axis the gas density  is parametrized by a Gaussian distribution with scale height $h$ ($=H/R$)

\begin{equation}
h = h_{\rm c} \Bigg( \frac{R}{R_{\rm c}}\Bigg)^{\psi} 
\end{equation}

\noindent
with $h_{\rm c}$ the critical scale height and $\psi$ the flaring angle. 
Following \citet{dAlessio06}, the settling of the large dust particles is implemented adopting two different power-law grain size populations, small 
(0.005 - 1\,\micron) and large (0.005 - 1000\,\micron) and power-law exponent $p=3.5$ (note that the size ranges are different from those adopted in \citealt{dAlessio06}). Dust mass absorption cross sections are from \citet{Andrews11}. The scale height of the two populations is $h$ for the small grains
and $\chi h$ ($\chi < 1$) for the large grains. The mass ratio between the two populations is controlled by the parameter $f_{\rm large}$: the dust surface density is $\Sigma_{\rm dust}  \ (1 - f_{\rm large})$  and $\Sigma_{\rm dust} \ f_{\rm large}$ for
the small and large grains, respectively. 
 
 \smallskip
 \noindent
 \textsc{dali} solves the 2D dust continuum radiative transfer and determines the dust temperature and
radiation field strength at each disk position. In a second step \textsc{dali} iteratively solves the gas thermal balance and chemistry, and returns as
output the continuum and line emission maps computed via ray tracing. Isotope selective photodissociation is included in the chemistry as described 
in \citet{Miotello14}.

\begin{figure*}[!t]
\centering
\includegraphics[width=19cm]{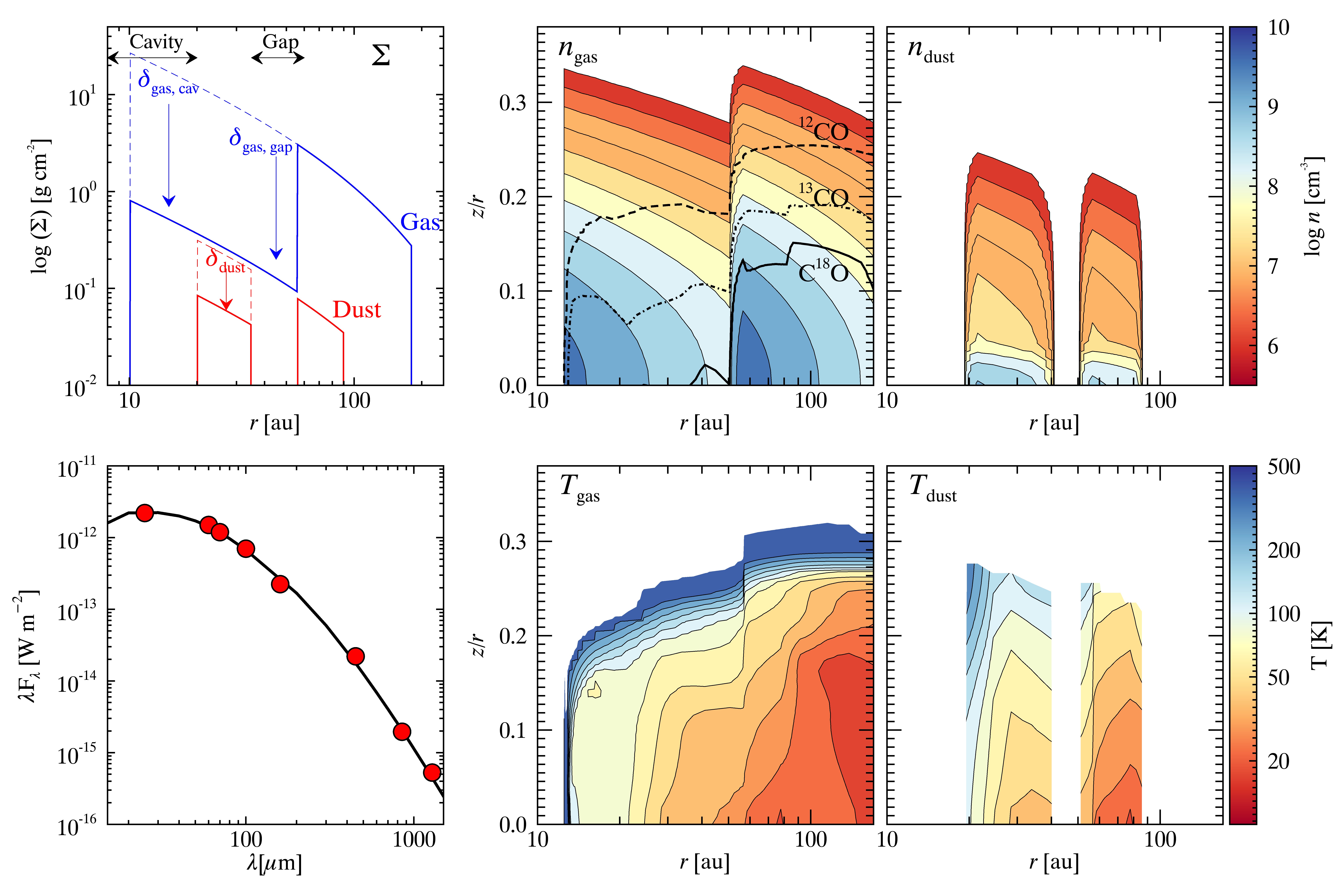}
\caption{Fiducial model:  ({\it top left}) surface density profile; ({\it top middle}) gas density structure: the $\tau = 1$ layer of the $J=2-1$ transition are
overlaid as dashed ($^{12}$CO), dot-dashed ($^{13}$CO) and solid (C$^{18}$O) curve; ({\it top right}) dust density structure;
({\it bottom left}) spectral energy distribution with fiducial model (black solid curve): data (red dots) are from IRAS, {\it Herschel} \citep{Pascual16}, SCUBA \citep{Sandell11} and ALMA (this work); ({\it bottom middle}) gas temperature structure; ({\it bottom right}) dust temperature structure.
}\label{fig:dali}
\end{figure*}

\subsection{Surface density}
Starting from the ALMA observations we define a surface density structure as shown in Figure~\ref{fig:dali} (left): the gas distribution extends from 
$R_{\rm gas \ in}$ to $R_{\rm gas \ out}$, while dust is only present between $R_{\rm dust \ in}$ -- $R_{\rm gap \ in}$ and between $R_{\rm gap \ out}$ -- $R_{\rm dust \ out}$. 
Along the radial axis, gas and dust densities are reduced by different decrease factors for the gas inside the dust cavity ($\delta_{\rm gas, cavity}$) and inside the dust gap
($\delta_{\rm gas, gap}$) and for the dust in the ring ($\delta_{\rm dust}$) as follow

\[ n_{\rm gas} =
  \begin{cases}
    0       & \quad \text{for } R \text{ $ < R_{\rm gas \ in}$}\\
     n_{\rm gas} \times \delta_{\rm gas, cavity}        & \quad \text{for }  \text{ $ R_{\rm gas \ in} < R < R_{\rm dust \ in} $}\\
     n_{\rm gas} \times \delta_{\rm gas, gap}           & \quad \text{for }  \text{ $ R_{\rm dust \ in} < R < R_{\rm gap \ out} $}\\
     n_{\rm gas}         & \quad \text{for }  \text{$R_{\rm gap \ out} < R < R_{\rm gas \ out}$}\\
     0                         & \quad \text{for }  \text{$R > R_{\rm gas \ out}$}\\
  \end{cases}
\]

\[ n_{\rm dust} =
  \begin{cases}
    0       & \quad \text{for } R \text{ $ < R_{\rm dust \ in}$}\\
     n_{\rm dust} \times \delta_{\rm dust}        & \quad \text{for }  \text{ $ R_{\rm dust \ in} < R < R_{\rm gap \ in} $}\\
     0                                                               & \quad \text{for }  \text{ $ R_{\rm gap \ in} < R < R_{\rm gap \ out} $}\\
     n_{\rm dust}         & \quad \text{for }  \text{ $ R_{\rm gap \ out} < R < R_{\rm dust \ out}$}\\
     0                           & \quad \text{for } \text{ $R > R_{\rm dust \ out}$}
  \end{cases}
\]

\noindent
Figure~\ref{fig:dali} shows the modelled disk density structure. 

\smallskip
\noindent
Note that our model does not include the small ring of hot dust, inside 0.27\,au \citep[e.g.,][]{Wagner15}. This does not influence our results as the extinction produced by this small amount of dust is negligible and it does not affect the propagation of the ultraviolet radiation further out in the disk.  

\subsection{Model grid}
We created a grid of disk models varying the most relevant geometrical and physical parameters 
These are:

\begin{itemize}
\item gas and dust mass (regulated by the combination of $\Sigma_{\rm c}$, $\gamma$, $R_{\rm c}$ and $\Delta_{\rm gd}$): which control the overall continuum and line emission as well as the SED
\item flaring angle ($\psi$): influencing mostly the radial intensity profile of the optically thick $^{12}$CO emission (arising from the outermost layers)
\item scale height ($h_{\rm c}$): which has a major impact on the intensity of the gas and dust emission and on the SED 
\item gas and dust decrease factors ($\delta_{\rm gas \ cavity}$, $\delta_{\rm gas \ gap}$, $\delta_{\rm dust}$): that control the gas and dust emission inside the dust gap 
\item size and position of the cavity and of the gap ($R_{\rm gas \ in}$, $R_{\rm dust \ in}$, $R_{\rm gap \ in}$, $R_{\rm gap \ out}$)
\item outer dust and gas radii ($R_{\rm dust \ out}$, $R_{\rm gas \ out}$)
\end{itemize}

\smallskip
\noindent
Table \ref{tab:dali} lists the definition, range and step size of the different parameters.

\noindent
Using \textsc{dali} we created model images of the dust continuum emission and of the $J=2-1$ transition of the three CO isotopologues.
In the case of the CO isotopologue lines, the model channel maps are computed with the same spectral resolution as the observations.
From the \textsc{dali} model images we measured synthetic visibilities, synthetic observations and residual (data-model) images 
reading the $uv$ coordinates, integration time, source position, hour angle and spectral window parameters directly from the observed ALMA 
measurement set. For this task we used the \textsc{casa} tools \textsc{simobserve} and \textsc{clean}. 

\smallskip
\noindent
The model grid is compared to the observations with the aim of constructing a fiducial model that quantitatively reproduces the ALMA observations. 
The observed radial intensity profiles (along the disk minor and major axes) and the SED are fitted against the model grid. 
The fiducial model is defined by the set of parameters that minimise the difference between observations and model grid within the explored parameter space.

\subsection{Fiducial model and comparison with observations}
\begin{figure*}[!t]
\centering
\includegraphics[width=18cm]{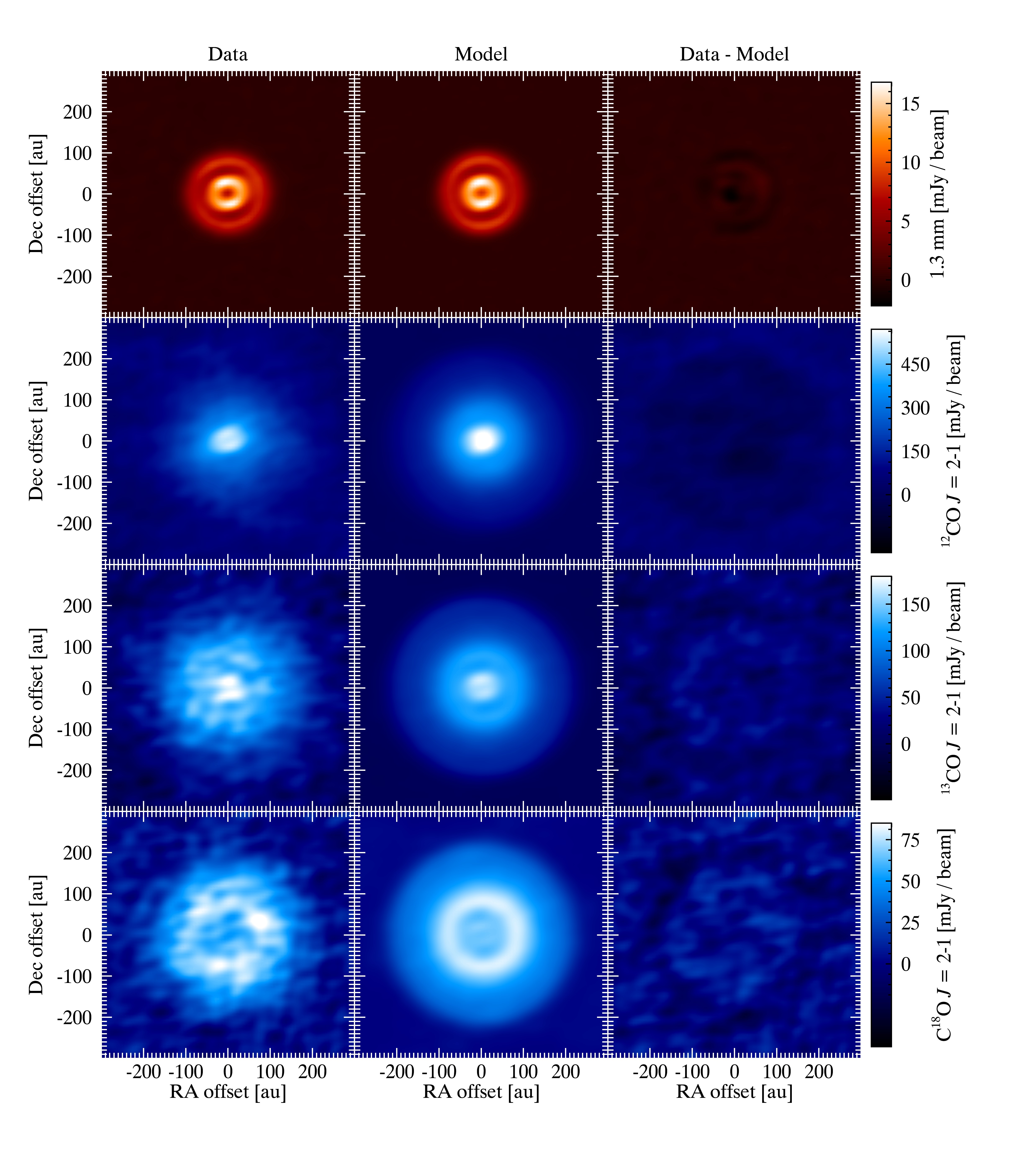}
\caption{Results of the fiducial model: (from left to right) ALMA image, model image, residual. Weighting scheme as in Figures~\ref{fig:cont} and ~\ref{fig:lines}.}\label{fig:model}
\end{figure*}

\begin{figure}
\centering
\includegraphics[width=9cm]{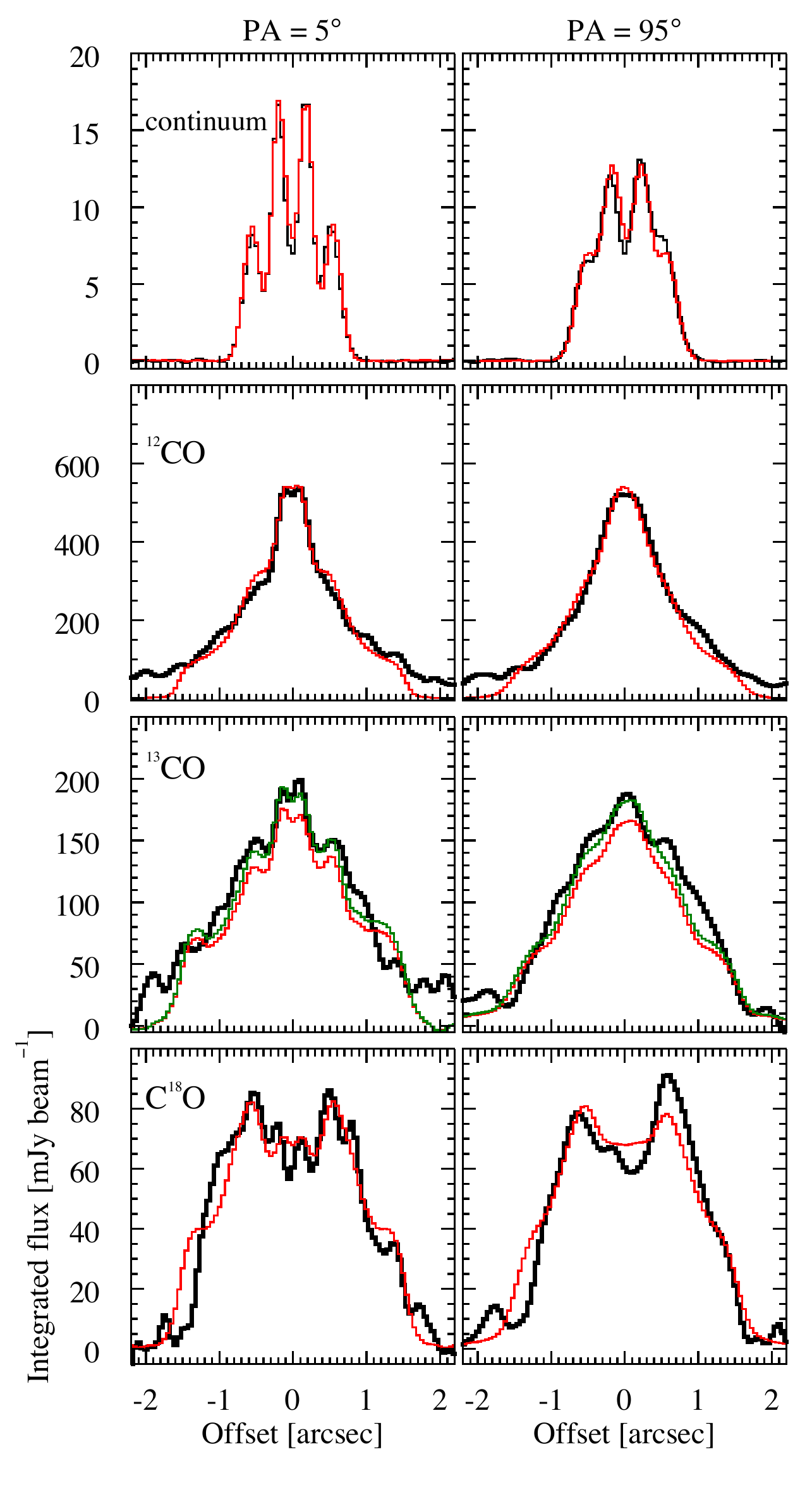}
\caption{Results of the fiducial model: comparison of the observed (black curve) and model (red curve) radial intensity profiles. In the case of $^{13}$CO
the green curve represents the fiducial model scaled up by 10$\%$ to match the absolute flux level.}\label{fig:radcut_model}
\end{figure}

The surface density distribution, the density and temperature structure and the SED of the fiducial model are shown in Figure~\ref{fig:dali}. 
Figures \ref{fig:model} and  \ref{fig:radcut_model} show the comparison with the observations: Figure~\ref{fig:model} 
shows (from left to right) the ALMA observations (continuum or line integrated map), synthetic observations and residual images (computed in the 
$uv$-plane); Figure~\ref{fig:radcut_model} shows the radial intensity profiles along the disk major and minor axes. In order to asses the quality of the 
fit, Figure~\ref{fig:grid} shows the radial profile differences for a subset of the model grid.

\smallskip
\noindent
{\em Dust surface density}: the fiducial model reproduces well the dust continuum image, with an inner dust cavity of $R \sim 20\,$au, an inner ring between 
$\sim 20 - 35\,$au, a gap between $\sim 35 - 56\,$au and an outer ring between $\sim 56 - 83\,$au. The cavity and the gap are empty of millimetre-sized 
dust particles down to the noise level; there is an upper limit to the dust mass of $\sim 10^{-2}$\,M$_{\earth}$ and $0.3$\,M$_{\earth}$ in the cavity and
gap, respectively. 
The inner ring is decreased by a factor of  $\sim 3.7$ (=$1/\delta_{\rm dust}$)  (Figure~\ref{fig:radcut}). The total dust mass is $1 \times 10^{-4}\,$M$_{\odot}$.
The cavity and the gap are also free of micron-sized dust particles as suggested by the NACO polarimetric observations (Figure~\ref{fig:radcut_naco}).

\smallskip
\noindent
{\em Gas in inner disk}: our analysis confirms the presence of gas inside the dust gap (i.e., 35\,au $< R <$ 56\,au) and inside the dust cavity ($R < 20$\,au)
down to an inner radius of $\sim 13\,$au. The gas surface density inward of the dust gap is decreased by a factor of $\sim$ 30 -- 40 
(=$1/\delta_{\rm gas \ gap}$). Interestingly we find that 
$\delta_{\rm gas \ cavity} = \delta_{\rm gas \ gap}$. 

\smallskip
\noindent
{\em Gas in the outer disk}: the gas surface density extends well beyond the dust outer radius. The fiducial model follows closely the slope of radial profile of 
the three CO isotopologues in the outer disk. The outer gas radius of 180\,au is set by the steep drop of the C$^{18}$O emission at large radii. Adopting a larger
outer gas radius would overestimate the C$^{18}$O emission in the outer disk, leaving almost unchanged the profiles of $^{12}$CO and C$^{13}$O.
The total gas mass is $1.9 \times 10^{-2}\,$M$_{\odot}$ for the adopted standard carbon abundance of [C]/[H]=$2.4 \times 10^{-4}$. 
While the fiducial model reproduces reasonably well the intensity profiles, the absolute flux of the $^{13}$CO emission is slightly underestimated. The difference in absolute flux is of the order of $\sim 10\%$. 

\section{Discussion}
The radial distribution of the (optically thin) 1.3\,mm emission showing two deep gaps is qualitatively consistent with the presence of multiple planets shaping
the dust distribution. \citet{Pinilla15} performed two dimensional hydrodynamical simulations coupled with a dust evolution model of a disk hosting two giant
planets showing that planets can produce multiple dust traps. These simulations show that the formation of dust traps is due to a combination of three main 
parameters: planetary mass, disk viscosity and dust fragmentation velocity.  In particular, the trap is easily formed for a high planetary mass and/or a low
 disk viscosity. 
%
%\smallskip
%\noindent
In the case of HD 169142, the lack of dust continuum emission inside the dust cavity ($R \lesssim 20\,$au) and in the outer gap 
(35\,au $\lesssim R \lesssim$ 56\,au) suggests very efficient dust trapping by means of two giant planets. 
%\citet{Reggiani14} estimate a mass of $\sim 30\,$M$_{\rm Jupiter}$ for the inner planet (inside the dust cavity). 
%Such a massive planet can easily filter out the dust particles giving rise to the inner dust cavity and to a dust trap outside 
%of the planet's orbit. 
%The density drop of gas inside the dust cavity of almost 2 orders of magnitudes requires a rather massive planet with 
%$M_{\rm planet} \gtrsim 1\, M_{\rm Jupiter}$. For lower planetary mass indeed the cavity is ``porous'' and gas from the outer disk could flow inwards filling
%the cavity  \citep[e.g.,][]{Lubow06, Alexander09}.

\smallskip
\noindent
Azimuthal asymmetries are sometimes observed in the dust continuum emission of Herbig AeBe systems as in the case of HD 142527 
\citep{Casassus12, Fukagawa13} and IRS 48 \citep{vdMarel13}. Such an asymmetry is thought to be due to dynamical interaction between the disk and a 
massive planet generating a vortex where dust particles are trapped \citep[Rossby instability, ][]{devalBorro07, Lyra09}. 
The lack of such an asymmetry in the two dust rings of HD 169142 implies an upper limit to the mass of the two planets of 
$\lesssim 10\,$M$_{\rm Jupiter}$.

\smallskip
\noindent
The density drop of gas inward of the dust gap requires a rather massive planet with M$_{\rm planet} \gtrsim 1\, $M$_{\rm Jupiter}$ for the outer planet. 
For lower planetary mass indeed the cavity is ``porous'' and gas from the outer disk could flow inwards filling the gap  \citep[e.g.,][]{Lubow06, Alexander09}.
This is consistent with the prescriptions of \citet{Rosotti16} which predict a mass of $\gtrsim 0.3\,$M$_{\rm Jupiter}$ (based on the size of the dust gap).

\smallskip
\noindent
In the case of the inner dust cavity, the ALMA CO maps show no further drop of the gas density, suggesting inward gas flow. This poses a further limit to the 
planetary mass inside the inner dust cavity of M$_{\rm planet} < $ M$_{\rm Jupiter}$.

\smallskip
\noindent
Thus, based on the lack of azimuthal asymmetric features and on the drop of the gas surface density, the ALMA data presented here are consistent 
with the presence of two giant planets of M$_{\rm planet} \sim 0.1 - 1\, $M$_{\rm Jupiter}$ and M$_{\rm planet} \sim 1-10\, $M$_{\rm Jupiter}$ for the inner and outer planet, respectively.

\smallskip
\noindent
The ALMA 1.3\,mm continuum image confirms the presence of a real dust gap (i.e., depletion of dust particles) between $\sim 35 - 56\,$au. This gap was
previously detected via near-infrared polarimetric imaging but only thanks to ALMA it is possible to determine the density drop. The dust gap is likely the
outcome of dynamical interaction between the disk and a second, unseen, planet as noted above. Moreover, while the dust continuum emission is confined
within a radius of 83\,au, with a sharp decay, the gas emission extends up to $\sim$ 180\,au radius as can be seen in Figures~\ref{fig:lines} and
\ref{fig:radcut}. This dichotomy is also observed in other systems seen at high angular resolution such as LkCa 15 \citep{Isella12}, HD 163296
 \citep{deGregorio13}, HD 100546 \citep{Walsh14}, HD 97048 \citep{Walsh16, vanderPlas16}. The smaller size of the dust disk compared to the gas is primarily due
    to an optical depth effect \citep[e.g.,][]{Dutrey97}, but the sharp drop of the dust emission however hints at the radial drift of the dust particles \citep{Weidenschilling77, Birnstiel14} 
with the grains trapped at local pressure maxima induced by the planet-disk interaction. 
The similarity of the radial intensity profile of the H-band scattered light and of the $^{12}$CO emission is indicative of dynamical and thermal coupling
of the small dust grains with the gas in outermost layers the disk.

\smallskip
\noindent
An alternative 
scenario that could also explain the surface density profile of HD 169142 is the case of magneto-rotational (MRI) instability creating dead-zones  
\citep[e.g.,][]{Regaly12, Flock15, Hasegawa15}. While a dead-zone itself results in a pressure bump with a gas density contrast of only a few, the combination 
with a mass loss mechanism can make this region largely devoid of gas 
by an amount similar to that 
observed in HD 169142. This process was investigated, e.g., by \citet{Morishima12} in the case of MRI instability combined with photoevaporation and 
by \citet{Pinilla16} including magnetohydrodynamics (MHD) wind (and dust evolution).

\smallskip
\noindent
Other protoplanetary systems around Herbig AeBe stars show similar multiple-ring dust distribution structures such as HD 100546 \citep{Walsh14}, HD 97048 \citep{Walsh16} and HD 163296 \citep{Isella16}. 
Also in these two cases, the ALMA observations suggest the existence of multiple planets which are responsible for the dynamical clearing of the disk. The emerging picture is that 
these Herbig AeBe protoplanetary disks are in an late evolutionary phase where planets have already formed at large distances from the star. 
Future ALMA observations will tell us whether this is true for the entire class of Herbig AeBe systems.

\begin{table*}[!t]
\centering
\caption{Fiducial disk model parameters}
\begin{tabular}{llll}
\hline
\hline
Parameter & \multicolumn{2}{c}{Value} & Description \\
\hline
& \multicolumn{2}{c}{fixed} &  \\
\hline
$M_{\star}$ [$M_{\odot}$] &  \multicolumn{2}{c}{1.65\tablefootmark{$\diamond$}} & stellar mass \\
T$_{\rm eff}$ [K]                & \multicolumn{2}{c}{8400\tablefootmark{$\ddagger$}} & stellar temperature \\
$L_{\star}$ [$L_{\odot}$]   & \multicolumn{2}{c}{10\tablefootmark{$\dagger$}} & stellar luminosity \\
d [pc]                                  & \multicolumn{2}{c}{117\tablefootmark{$\dagger$}} & stellar distance \\
$i$ [$^{\circ}$]                   &  \multicolumn{2}{c}{13\tablefootmark{$\star$, $\dagger$}} & disk inclination  \\
PA [$^{\circ}$]                    & \multicolumn{2}{c}{5\tablefootmark{$\star$, $\dagger$}} & disk position angle \\ 
$\chi$, f$_{\rm large}$ & \multicolumn{2}{c}{0.2\tablefootmark{$\dagger$}} & settling parameters\\
\hline
& range / step & best fit & \\
\hline
$\psi$                                & 0.0 -- 0.2 / 0.05 & 0.0 & degree of flaring \\
$\gamma$                        &  0 -- 2 / 0.5 & 1.0 & $\Sigma (r)$ power-law exponent \\
$h_{\rm c}$                       & 0.05 -- 0.09 / 0.01 & 0.07 & scale height at $R = R_{\rm c}$\\
$R_{\rm c}$  [au]              & 50 -- 200 / 50 & 100 & critical radius  \\
$\Delta_{\rm gd}$             & 50 -- 100 / 10 & 80 & gas-to-dust mass ratio \\
$\Sigma_{\rm c}$ [g\,cm$^{-2}$]    &  5 -- 8 / 0.5 & 6.5 & $\Sigma_{\rm gas}(R)$ at $R=R_{\rm c}$\\
$R_{\rm gas \ in}$  [au] &  11 -- 15 / 1 & 13 & gas inner radius \\
$R_{\rm dust \ in}$  [au] & 18 -- 22 / 1 &  20 & dust inner radius \\
$R_{\rm gap \ in}$ [au]  & 32 -- 36 / 1 & 35 & dust cavity inner radius \\
$R_{\rm gap \ out}$ [au] & 52 -- 60 / 2 & 56 & dust outer radius \\
$R_{\rm gas \ out}$ [au] & 170 -- 210 / 10 & 180 & gas outer radius\\
$R_{\rm dust \ out}$ [au] & 81 -- 91 / 2 & 83 & dust outer radius\\
$\delta_{\rm dust}$            &  0.25 -- 0.35  /   0.02    & 0.27  & dust depletion for $R_{\rm dust \ in} < R < R_{\rm gap \ in}$ \\
$\delta_{\rm gas \ cavity}$ &  0.01 -- 0.05  /   0.005  & 0.025 & gas depletion for $R_{\rm gas \ in} < R < R_{\rm dust \ in}$ \\
$\delta_{\rm gas \ gap}$    &  0.01 -- 0.05  /   0.005  & 0.025 & gas depletion for $R_{\rm dust \ in} < R < R_{\rm gap \ in}$ \\
\hline
\hline
\end{tabular}\label{tab:dali}
\tablefoot{
References: \tablefoottext{$\diamond$}{\citet{Blondel06}}; \tablefoottext{$\ddagger$}{\citet{Dunkin97}}; \tablefoottext{$\star$}{\citet{Raman06}; \tablefoottext{$\dagger$}{this work}}.
}
\end{table*}

\section{Conclusions}
From the analysis of the ALMA observations presented in this paper we conclude that gas and dust in HD 169142 are physically decoupled with the 
dust particles concentrated in two rings between $\sim 20 - 35$\,au and $\sim 56-83$\,au. Thanks to ALMA, we have, for the first time, strong constraints 
on the distribution and dynamics of bulk of the dust and gas. We find a real dust depletion (i.e., absence of dust particles) inside the cavity 
($R \lesssim 20\,$au) and the gap ($35 \lesssim R \lesssim 56 $\,au). The dust cavity and gap are filled in with gas as suggested by the emission maps of the three CO isotopologues, with the gas surface density reduced by a factor of $\sim 30-40$ for $R \lesssim 56\,$au. The sharp edge of the continuum map at 83\,au is indicative 
of radial drift of dust grains (dust outer radius $<<$ gas outer radius).

\smallskip
\noindent
Among the various theories proposed to explain the opening of gaps in disks (e.g., dynamical interaction with planets, MRI, dust sintering and 
photoevaporation), the most likely scenario for HD 169142 is the presence of multiple giant planets ($\gtrsim $M$_{\rm Jupiter}$) carving out the disk and giving
rise to the cavity and the gap and trapping the dust particles beyond the planetary orbits. The combined effect of MRI instability (forming dead-zones)
and MHD wind could also give rise to the rings and gaps structure in both gas and dust although with different characteristics that could be tested by future higher S/N and higher angular resolution data.

\smallskip
\noindent
We stress the importance of spatially resolved observations of multiple CO isotopologues transitions: thanks to the different optical depths, the spatially
resolved channel maps of the three isotopologues allows us to detect and quantify the very small amount ($\lesssim 0.1\,$M$_{\rm Jupiter}$) of gas inside 
the dust gap. This demonstrates the potential of CO isotopologues observations in probing the gas surface density and the evolutionary phase of
protoplanetary systems.

\begin{acknowledgements}
This paper makes use of the following ALMA data: ADS$/$JAO.ALMA$\#$2013.1.00592.S. ALMA is a partnership of ESO (representing its member states), NSF (USA) and NINS
(Japan), together with NRC (Canada), NSC and ASIAA (Taiwan), and KASI (Republic of Korea), in cooperation with the Republic of Chile. The Joint ALMA Observatory is operated by 
ESO, AUI$/$NRAO and NAOJ.
This work has made use of data from the European Space Agency (ESA)
mission {\it Gaia} (\url{http://www.cosmos.esa.int/gaia}), processed by
the {\it Gaia} Data Processing and Analysis Consortium (DPAC,
\url{http://www.cosmos.esa.int/web/gaia/dpac/consortium}). Funding
for the DPAC has been provided by national institutions, in particular
the institutions participating in the {\it Gaia} Multilateral Agreement.
DF acknowledges support from the Italian Ministry of Education, Universities and Research, project SIR (RBSI14ZRHR). 
MC and MRH are supported by a TOP grant from the Netherlands Organisation for Scientific Research (NWO, 614.001.352).
CW acknowledges financial support from the Netherlands Organisation for Scientific Research (NWO, grant 639.041.335) and start-up funds from the University of Leeds, UK. 
The authors thank M. Tazzari, S. Facchini, G. Rosotti, L. Testi and P. Pinilla for useful
 discussions and S. Quanz for providing the NaCo image.
\end{acknowledgements}

%\bibliographystyle{aa} % style aa.bst
 %\bibliography{mybib} % your references Yourfile.bib
%

\begin{appendix}

\section{Channel maps}
\begin{figure*}[!t]
\centering
\includegraphics[width=18cm]{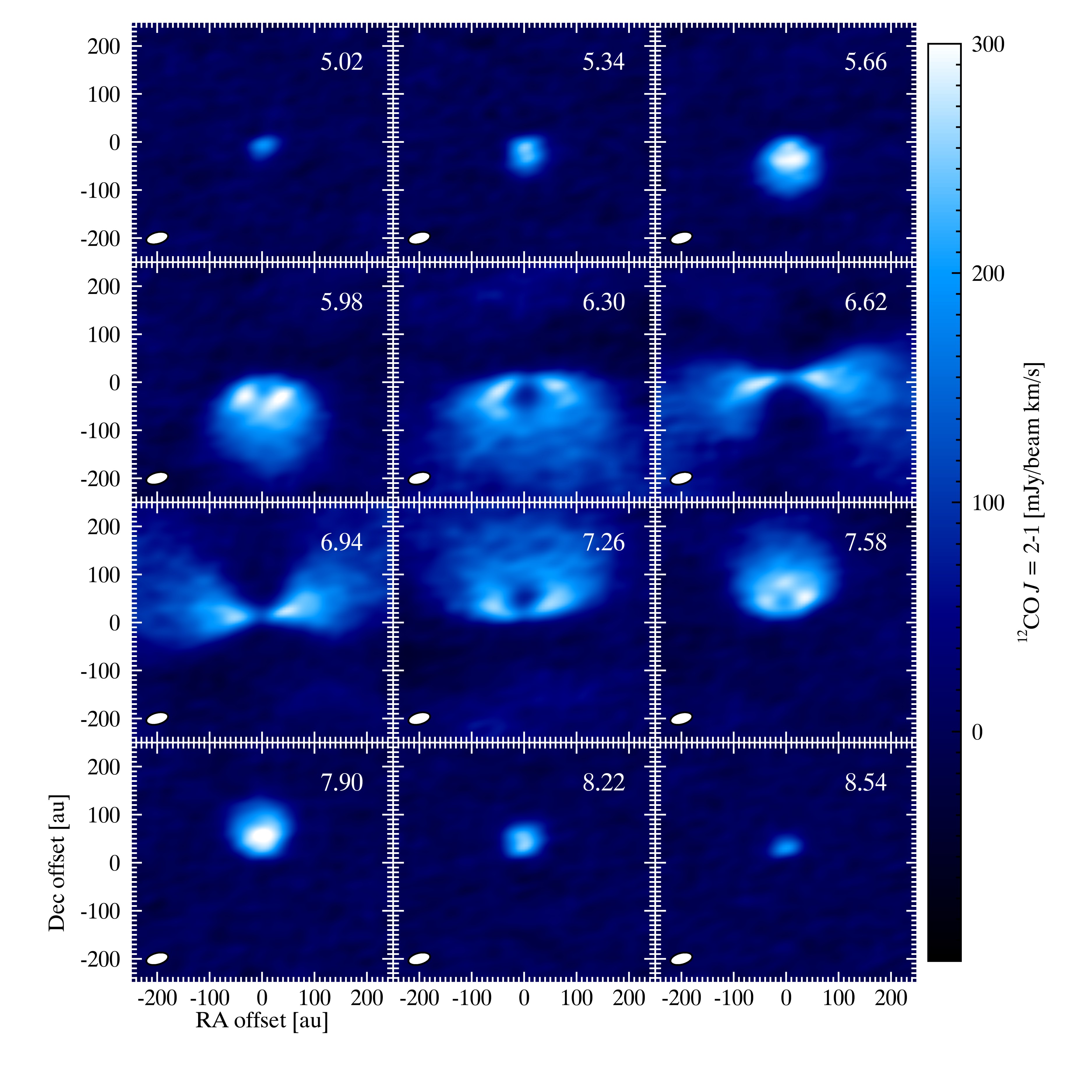}
\caption{ALMA channels maps of $^{12}$CO $J=2-1$.}\label{fig:channels_12co}
\end{figure*}

\begin{figure*}[!t]
\centering
\includegraphics[width=18cm]{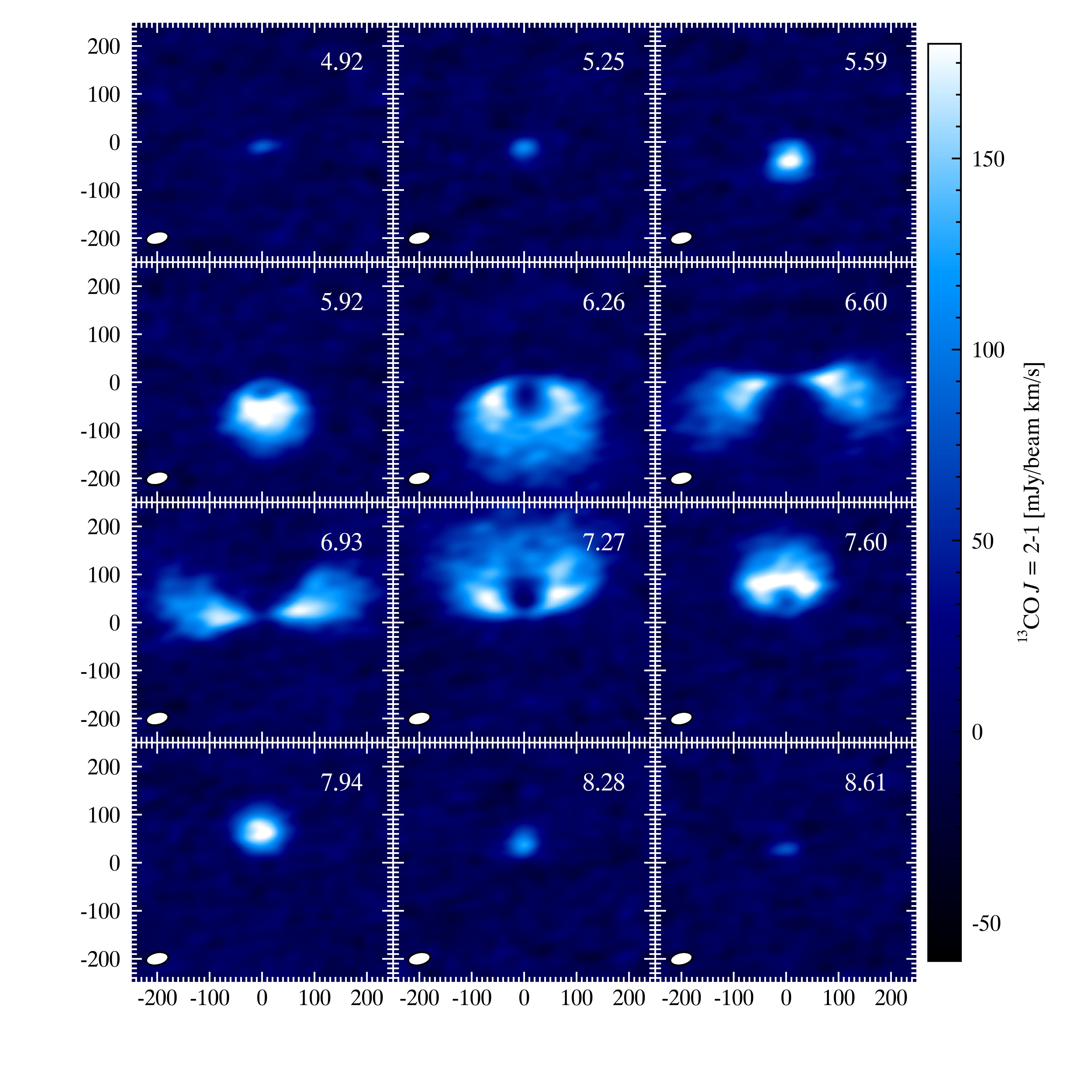}
\caption{ALMA channels maps of $^{13}$CO $J=2-1$.} \label{fig:channels_13co}
\end{figure*}

\begin{figure*}[!t]
\centering
\includegraphics[width=18cm]{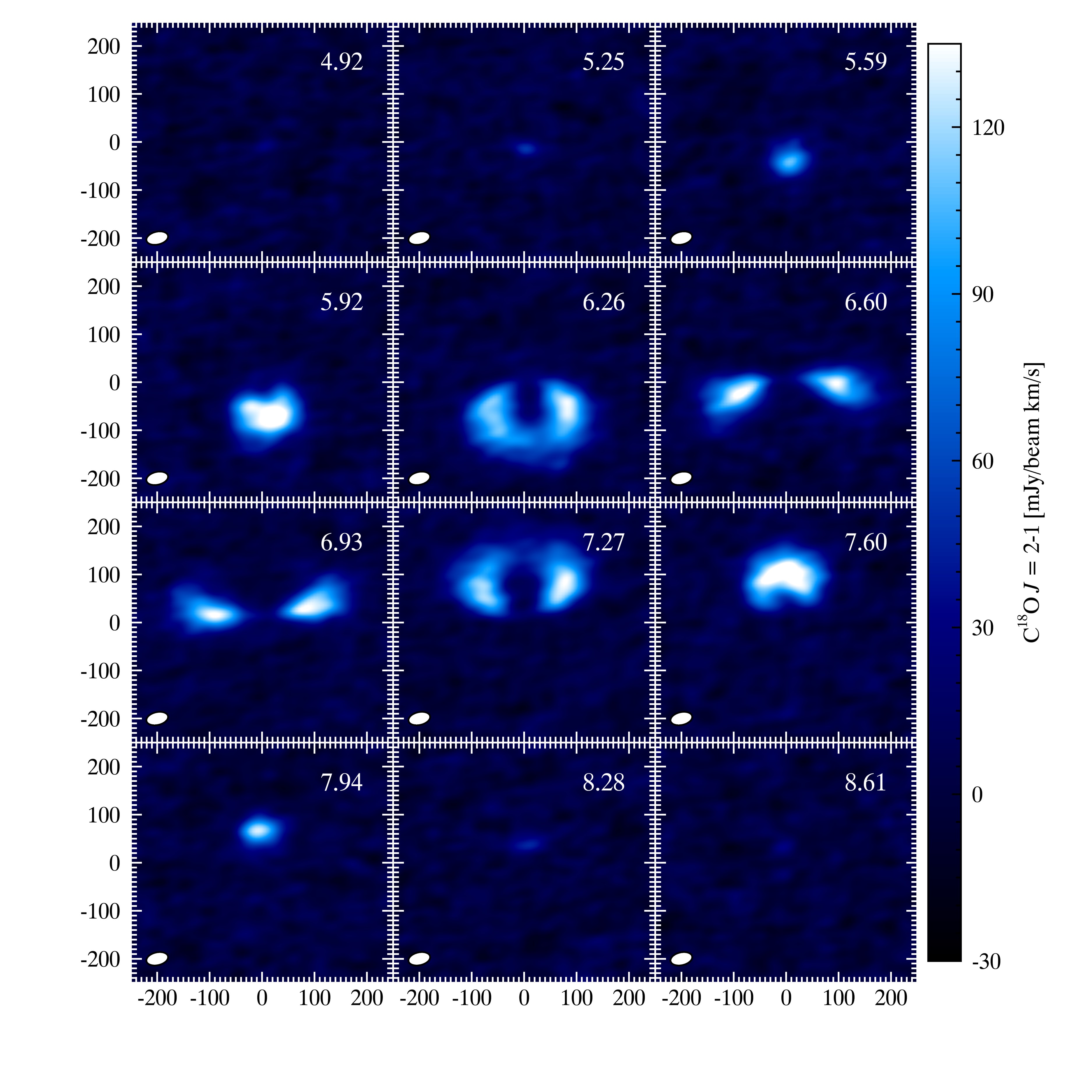}
\caption{ALMA channels maps of C$^{18}$O $J=2-1$.}\label{fig:channels_c18o}
\end{figure*}

The individual channel maps of $^{12}$CO, $^{13}$CO and C$^{18}$O are shown in Figures~\ref{fig:channels_12co}, ~\ref{fig:channels_13co} and ~\ref{fig:channels_c18o}, 
respectively. The velocity scale is defined in the local standard of rest (LSR) system. In all cases, the maps are created using natural weighting and the velocity resolution is resampled to 0.160\,\kms
 ($^{12}$CO) and 0.168\,\kms ($^{13}$CO and C$^{18}$O) with Hanning smoothing. The disk emission is detected from $v_{\rm LSR} = 4.9\,$\kms to $v_{\rm LSR}=8.6\,$\kms
 and the systemic velocity is $v_{\rm LSR}=6.9\,$\kms.

\section{Model grid}
\begin{figure*}
\includegraphics[width=18cm]{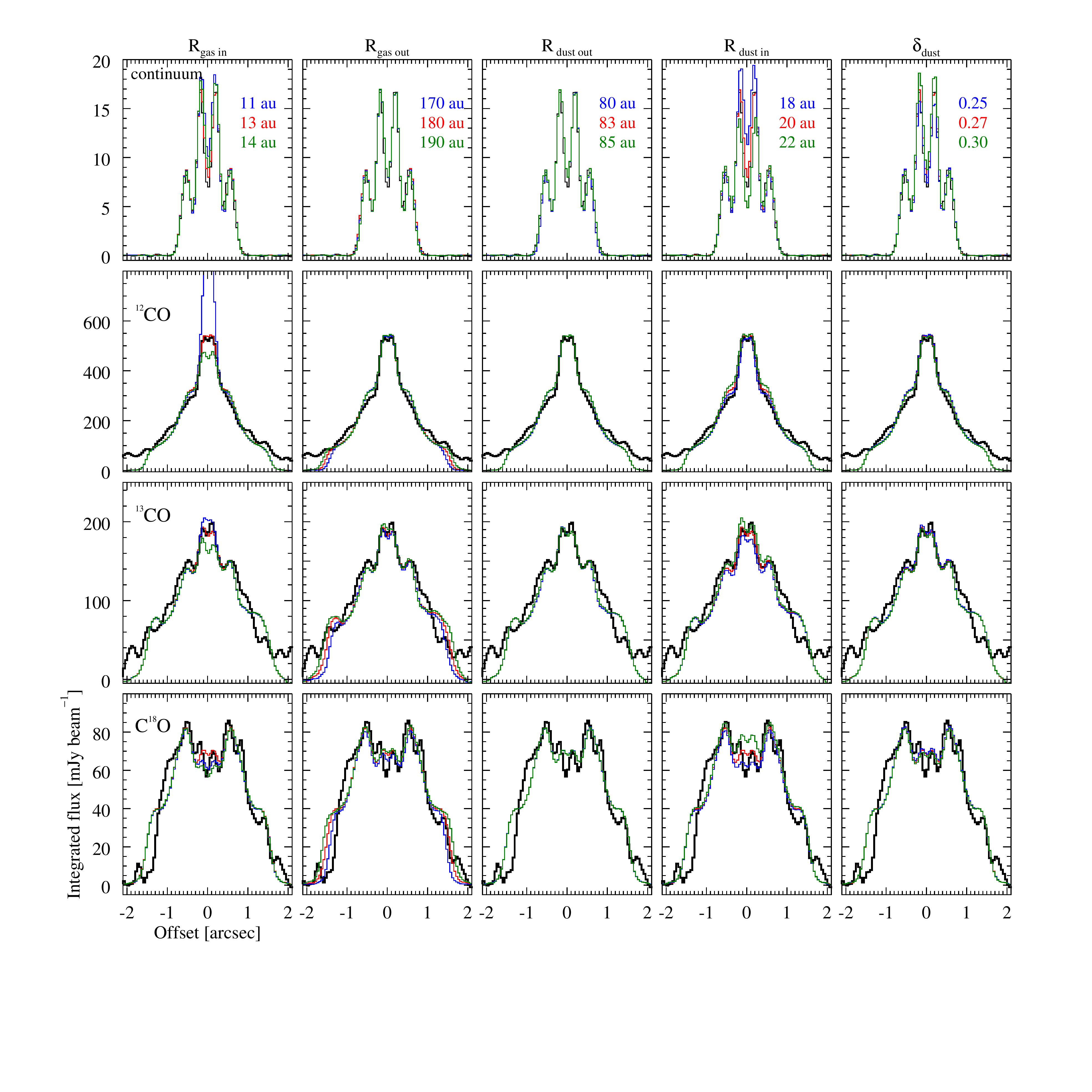}
\caption{Radial intensity profile differences (PA=5$^{\circ}$) for a subset of the model grid. The $^{13}$CO models are scaled up by 10\% to match
the absolute flux level of the observations. }\label{fig:grid}
\end{figure*}

Figure \ref{fig:grid} shows the results of the model grid (sed. 5.3) where we vary one parameter at a time. Only a subset of the model grid is shown here.
Note in particular, how the $^{12}$CO radial profile is very sensitive to $R_{\rm gas \ in}$ (the inner gas radius) while the inner dust radius ($R_{\rm cav}$)
affects not only the continuum radial profile but also the strength of the $^{13}$CO and C$^{18}$O emission in the inner disk.  

\end{appendix}

\end{document}